# Supernovae in paired host galaxies

T.A. Nazaryan[1], A.R. Petrosian[1], A.A. Hakobyan[1], V.Zh. Adibekyan[2], D. Kunth[3], G.A. Mamon[3], M. Turatto[4], L.S. Aramyan[1]

[1]Byurakan Astrophysical Observatory, Armenia, e-mail: nazaryan@bao.sci.am
[2]Centro de Astrofísica da Universidade do Porto, Portugal
[3]Institut d'Astrophysique de Paris, France
[4]INAF-Osservatorio Astronomico di Padova, Italy

*Abstract.* We investigate the influence of close neighbor galaxies on the properties of supernovae (SNe) and their host galaxies using 56 SNe located in pairs of galaxies with different levels of star formation (SF) and nuclear activity. The mean distance of type II SNe from nuclei of hosts is greater by about a factor of 2 than that of type Ibc SNe. For the first time it is shown that SNe Ibc are located in pairs with significantly smaller difference of radial velocities between components than pairs containing SNe Ia and II. We consider this as a result of higher star formation rate (SFR) of these closer systems of galaxies. SN types are not correlated with the luminosity ratio of host and neighbor galaxies in pairs. The orientation of SNe with respect to the preferred direction toward neighbor galaxy is found to be isotropic and independent of kinematical properties of the galaxy pair.

*1. Introduction.* According to large-sample statistical studies, in most cases gravitational interaction can be a triggering mechanism for nuclear activity and/or circumnuclear starburst in interacting and merging galaxies [1,2]. Observational results suggest that core-collapse (CC) SNe are tightly connected with recent SF, e.g. [3,4,5]. The radial distribution of SNe, especially CC SNe in active and star-forming galaxies shows a higher concentration toward the center of the active hosts than in normal ones [6]. The fraction of type Ibc SNe in central regions of disturbed hosts is higher compared to that of the undisturbed hosts [7]. There is also an indication that SN rates are higher in galaxy pairs compared with that in groups, which can be related to the enhanced SFR in strongly interacting systems [8].

The aim of this study is to investigate to what extent gravitational interaction with a close neighbor can be connected with nuclear activity and/or enhanced SF in galaxy pairs, using SNe as tracers of recent SF. We selected samples of paired galaxies with different nuclear/starforming activity levels, from AGNs and starburst galaxies to passive ones. We examined correlations between kinematical properties of pairs of galaxies, integral parameters of SN hosts and their neighbors, as well as SN types and their distributions. The complete study is presented in paper [9].

*2. Sample.* The sample of the current study was obtained by cross-matching the catalog of SNe by [10] with the sample of selected pairs of galaxies (see below). SN database [10] contains 3876 SNe located in the area of the sky covered by the Sloan Digital Sky Survey (SDSS) Data Release 8 (DR8). We used three catalogs of galaxies with different levels of nuclear activity to construct our sample of close pairs of galaxies. These catalogs are the following: (1) the catalog of Markarian (MRK) galaxies, (2) the Second Byurakan Survey (SBS) galaxies catalog, and (3) the North Galactic Pole (NGP) galaxy catalog.

Results of a close neighbors search for MRK galaxies within position-redshift space are published in [11]. Three criteria for redshifts and pair separation were used to select the sample of close neighbors of MRK galaxies in [11]. We also conducted the search of neighbors for SBS and NGP galaxies using the same criteria. The sample of 675 pairs of galaxies was cross-matched with the list of SN hosts from [10]. In total 56 SNe in 44 hosts were identified. Radial distances of SNe from their host nuclei are normalized to $R_{25}$ of hosts according to [5]. Pairs of galaxies are described via two parameters describing strength/stage of interacting/merging: difference of radial velocities $\Delta v_r$ of SN hosts and their neighbors and linear projected distance $D_p$ between pair members.

*3. Statistics and discussions.* The first row of Table 1 shows mean morphological *t*-types of our paired sample hosts for SNe of different types. Hosts of type Ibc and II SNe are of later morphological classes than those of type Ia. This is well known observational result, e.g. [3]. For comparison, in the second row, mean morphologies of the unbiased sample of 1021 nearby hosts of different types SNe from [10] are presented. There is no statistically significant difference between our and [10] *t*-types. SN hosts of our sample tend to form pairs with neighbors with similar morphologies. The percentage of barred galaxies among our hosts is larger (2σ) than that in the sample from [10]. Since our sample consists of paired hosts only, the excess of bars is expected.

*Table 1.* Morphological classification of SN hosts and their neighbors.

| | | All SNe | Ia | Ibc | II |
|---|---|---|---|---|---|
| 1 | Our host mean *t*-type | 3.4±0.4 | 2.0±0.6 | 4.2±0.7 | 5.3±0.7 |
| 2 | Host sample from [10], mean *t*-type | 3.9±0.1 | 2.3±0.2 | 4.5±0.2 | 5.0±0.1 |
| 3 | Our neighbor mean *t*-type | 4.1±0.5 | 2.5±0.5 | 5.9±0.9 | 4.7±1.1 |
| 4 | *p* value KS test for rows 1 & 2 | 0.43 | 0.74 | 0.43 | 0.89 |
| 5 | *p* value KS test for rows 1 & 3 | 0.00 | 0.14 | 0.71 | 0.46 |
| 6 | Our host, bar (%) | 45±7 | 42±12 | 50±15 | 47±13 |
| 7 | Host sample from [10], bar (%) | 30±1 | 27±3 | 28±4 | 31±2 |
| 8 | Our neighbor, bar (%) | 32±6 | 26±10 | 42±15 | 33±13 |

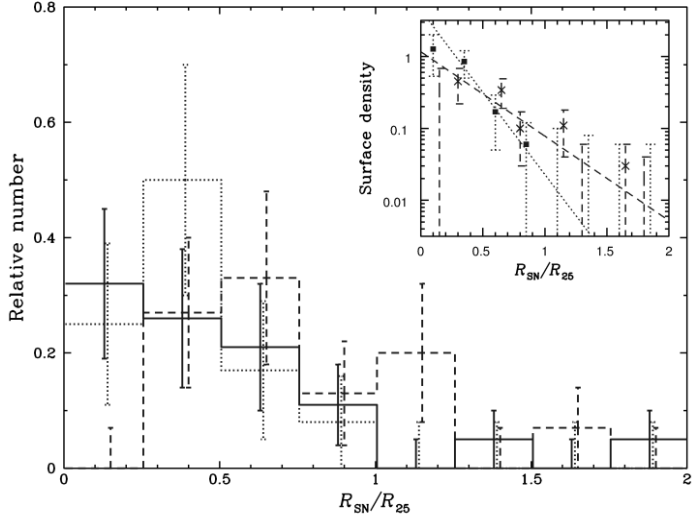

***Figure 1.*** Normalized histograms of radial distributions of types Ia (*solid*), Ibc (*dotted*), and II (*dashed*) SNe. The *error bars* assume a Poisson distribution. *Top-right*: surface density profiles (with arbitrary normalization) of all CC SNe: Ibc (*dotted*), and II (*dashed*). The *lines* show the maximum likelihood exponential surface density profiles of CC SNe.

The radial distribution of SNe is shown in Fig. 1. The mean value of $R_{SN}/R_{25}$ is 0.53±0.10 for type Ia. The mean normalized distance $R_{SN}/R_{25}$ of SNe II, at 0.74±0.09 is roughly double that of SNe Ibc at 0.38±0.06. The significance of more concentrated Ibc SNe relative to II SNe is 3.35σ. Our mean values for CC SNe within estimated errors are in agreement with those of [5]. The higher II to Ibc ratio of mean normalized distances in our sample is marginally significant (1.54σ) than that in the [5] sample.

In addition, we calculated surface densities of type Ibc and II SNe, assuming that they are located within discs of spiral hosts, as is shown in top-right corner of Fig. 1. To check consistency of surface density distributions of CC SNe with an exponential model, we generated exponential distributions with the corresponding scale lengths using maximum likelihood fitting according to [5] and compared real SNe distributions with them. The observational deficit of SNe in the central regions due to the Shaw effect is the main source of the deviation. There is non-significant excess of Ibc SNe in central regions of hosts of our sample in comparison with those of [5] sample.

We analyzed subsamples of pairs with SNe of different types, with the main results presented in Table 2. The distribution of $\Delta v_r$ of pairs containing SNe Ia and II is the same with practically consistent mean values. However, the same distributions of pairs with Ibc and II SNe are significantly different (3.2σ) with smaller mean of $\Delta v_r$ for Ibc. This means that Ibc SNe explode preferably in pairs with stronger interaction. However, there is no significant difference between mean values of $D_p$ of pairs with SNe of different types as is seen in Table 2.

***Table 2.*** Parameters of pairs containing SNe of different types.

| SN type | Number of SNe | Mean $\Delta v_r$ (km s$^{-1}$) | Mean $D_p$ (kpc) |
|---|---|---|---|
| Ia | 19 | 134±27 | 34±4 |
| Ibc | 12 | 56±14 | 28±5 |
| II | 15 | 129±18 | 32±6 |
| All | 56 | 117±9 | 33±3 |

To explain the strong dependence of CC SNe types on $\Delta v_r$, we considered SFR as the main parameter, affecting SN production in galaxies. SN rate and number ratio of type Ibc to type II SNe should increases with increasing of SFR [3,4,6]. Thus, we expect a relatively larger amount of star-forming galaxies, especially with smaller $\Delta v_r$ and $D_p$ in our sample due to interaction-triggered starbursts [1,2]. Therefore, excess of Ibc SNe compared to II SNe in the pairs with smaller $\Delta v_r$ and $D_p$ can be a result of higher SFR.

As a conclusion, we consider that close environment of galaxies can have some observable effect on SN production due to the impact on SF of galaxies.

A.R.P., A.A.H., and L.S.A. are supported by the Collaborative Bilateral Research Project of the State Committee of Science (SCS) of the Republic of Armenia and the French Centre National de la Recherché Scientifique (CNRS). This work was made possible in part by a research grant from the Armenian National Science and Education Fund (ANSEF) based in New York, USA. V.Zh.A. is supported by grant SFRH/BPD/70574/2010 from FCT (Portugal) and would further like to thank for the support by the ERC under the FP7/EC through a Starting Grant agreement number 239953.